\documentclass[prd,amsmath,amssymb]{revtex4}

\usepackage{graphicx}
\usepackage{dcolumn}
\usepackage{bm}
\usepackage{cases}

\begin{document}

\title{Symmetries of the Burgers Turbulence without Pressure \\ }

\author{Timo\hspace*{1mm}Aukusti\hspace*{1mm}Laine \vspace*{0.5cm}}
\email{timo.au.laine@wmail.fi}



           
\begin{abstract}
We investigate local symmetries of the Burgers turbulence driven by an external random force. By using a path integral formalism, we show that the Jacobian has physics in it; local symmetries and an anomaly. We also study a local invariance of the effective action and show it is related to Kolmogorov's second law of self-similarity.

\end{abstract}

\maketitle

\vspace*{0.5cm}

\section{\label{sec1}Burgers equation}

In this paper we investigate one-dimensional Burgers equation 

\begin{equation}
  u_t + uu_x -\nu u_{xx} = f(t,x), \label{eq1}
\end{equation}

\noindent
which is driven by Gaussian random force $f(t,x)$,

\begin{equation}
  \langle f(t,x)f(t',y) \rangle = \kappa(x-y)\delta(t-t'). \label{eq2}
\end{equation}

\noindent
Function $\kappa$ defines the spatial correlation of the random forces and $\nu$ is the viscosity. The equation has been studied in several papers, see for example Refs.~\onlinecite{polyakov}-\onlinecite{ivaskevich}.

We use the path integral formalism and write equations (\ref{eq1})-(\ref{eq2}) as

\begin{equation}
  \langle F[\lambda]\rangle = \int D\mu Du F[\lambda] J[u]\exp(-S[u,\mu]).
\end{equation}

\noindent
Action $S$ is defined as

\begin{equation}
  S[u,\mu] = \frac{1}{2} \int dtdxdy \mu(t,x)\kappa(x-y)\mu(t,y) -i\int
dtdx \mu(u_t+uu_x-\nu u_{xx}),
\end{equation}

\noindent
and Jacobian is

\begin{equation}
  J[u] = \det \biggl | \frac{\delta f}{\delta u}\biggr | = \det | \partial_t+u_x+u\partial_x-\nu \partial_{xx} |.
\label{eq:jac}
\end{equation}

We also use the following representation for the determinant,

\begin{equation}
  J[u] = \int D\bar{\Psi}D\Psi \exp(-S_A),
\end{equation}

\noindent
where action  is

\begin{equation}
  S_A = -\int dtdx \bar{\Psi}(\partial_t+u_x+u\partial_x-\nu \partial_{xx})\Psi. \label{det1}
\end{equation}

\noindent
Fields $\Psi = \Psi(t,x,u)$ and $\tilde{\Psi} = \tilde{\Psi}(t,x,u)$ are anticommuting functions, Refs.~\onlinecite{zinn}-\onlinecite{ramond}.

\section{\label{sec2}Determinant}

In this section we study local symmetries of determinant action (\ref{det1}) and calculate the Jacobian.  When the viscosity is set to zero, action (\ref{det1}) is  invariant under  local time reparametrization, $\alpha = \alpha(t)$,

\begin{eqnarray}
  &&\delta u_{A1} = \alpha u_t+ \alpha'u, \label{detu} \\
  &&\delta \Psi_{A1} = \alpha \Psi_t,\\
  &&\delta \bar{\Psi}_{A1} = \alpha \bar{\Psi}_t.
\end{eqnarray} 

\noindent
Action (\ref{det1}) is also invariant under local space-time symmetry, $\epsilon = \epsilon(t,x)$, $\nu = 0$,

\begin{eqnarray}
  &&\delta u_{A2} = \epsilon u_x-\epsilon_x u -\epsilon_t, \\
  &&\delta \Psi_{A2} = \epsilon \Psi_x+\epsilon_x\Psi, \\
  &&\delta \bar{\Psi}_{A2} = \epsilon \bar{\Psi}_x.\label{detpsi}
\end{eqnarray} 

Variation with respect to $u$ gives, $\nu = 0$,

\begin{equation}
  \frac{\delta S_A}{\delta u} =  \bar{\Psi}_x\Psi - \frac{\delta ( \bar{\Psi}\Psi_t) }{\delta u}+ u\frac{\delta ( \bar{\Psi}_x\Psi )}{\delta u}. \label{det22}
\end{equation}

\noindent
If assuming that fields $\Psi$ and $\bar{\Psi}$ are independed on $u$, variation (\ref{det22}) is still nonvanishing.  Alternatively, if variation (\ref{det22}) is claimed to be zero,  field $u$ has a constraint. This shows that action (\ref{det1}) depends on field $u$ and determinant (\ref{eq:jac}) is field $u$ dependent. The problem here is how to calculate the determinant in such a manner that the causality is preserved. 

The reason for the ambiguity is that in the current situation operator in $\delta f/\delta u$ is non-self-adjoint. According to the definition $\det | \delta f/\delta u|$ is the product of the eigenvalues $\delta f/\delta u$. A way to calculate the determinant is to consider the following eigenvalue equations,

\begin{numcases}{}
(\partial_t+u_x+u\partial_x -  \nu \partial_{xx})A=\lambda A, \\ 
(-\partial_t-u\partial_x -  \nu \partial_{xx})A=\lambda A, 
\end{numcases}

\noindent
where $\lambda = \lambda(t,x,u)$ is an eigenvalue and $A = A(t,x,u)$ is the corresponding eigenfunction. This gives a result 

\begin{equation}
  \lambda = \frac{1}{2}u_x - \nu \frac{\partial_{xx}A}{A} = \frac{1}{2}u_x - \nu G[u],
\end{equation} 

\noindent
and $G[u]$ is an $u$-dependent function.
When using the result, the full determinant action takes the form 

\begin{equation}
   S_{D} = -  \int dtdx \bar{\Psi}(\frac{1}{2}u_x - \nu G[u])\Psi  , \label{det2}
\end{equation}

\noindent
or

\begin{equation}
    J[u] = \exp \Bigl (  \int dtdx \ln ( \frac{1}{2} u_x - \nu G[u]) \Bigr ) =  \exp (-\bar{S}_{D}). \label{det20}
\end{equation}

\noindent 
 The effective action is then

\begin{equation}
  S_{eff}[u,\mu] =  \frac{1}{2} \int dtdxdy \mu(t,x)\kappa(x-y)\mu(t,y) -i\int
dtdx \mu(u_t+uu_x-\nu u_{xx}) -   \int dtdx \ln( \frac{1}{2} u_x-\nu G[u]). \label{acteff}
\end{equation}

\noindent
We have therefore shown that generally Burgers equation has a non-zero Jacobian. 

\section{\label{sec3}Anomaly }

We take a look of action (\ref{det2}). It can be written as

\begin{eqnarray}
   S_{D} &=&  \frac{1}{2}  \int dtdx u(\bar{\Psi}_x\Psi + \bar{\Psi}\Psi_x)+ \nu  \int dtdx G[u]\bar{\Psi}\Psi  , \label{det021}\\
              &=& \frac{S_A+S_B}{2} \label{det022},
\end{eqnarray}

\noindent
where 

\begin{equation}
  S_B = -\int dtdx \bar{\Psi}(-\partial_t-u\partial_x-\nu \partial_{xx})\Psi. \label{det01}
\end{equation}

\noindent
The full determinant action (\ref{det021}) is therefore a sum of two non-self-adjoint operators $S_A$ and $S_B$.

Action (\ref{det01}) is invariant under local time reparametrization, $\alpha = \alpha(t)$,

\begin{eqnarray}
  &&\delta u_{B1} = \alpha u_t+ \alpha'u, \label{0detu} \\
  &&\delta \Psi_{B1} = \alpha \Psi_t,\\
  &&\delta \bar{\Psi}_{B1} = \alpha \bar{\Psi}_t,
\end{eqnarray} 

\noindent
and under local space-time symmetry, $\epsilon = \epsilon(t,x)$, $\nu = 0$,

\begin{eqnarray}
  &&\delta u_{B2} = \epsilon u_x-\epsilon_x u +\epsilon_t, \\
  &&\delta \Psi_{B2} = \epsilon \Psi_x, \\
  &&\delta \bar{\Psi}_{B2} = \epsilon \bar{\Psi}_x+\epsilon_x\bar{\Psi}.\label{0detpsi}
\end{eqnarray} 

Now we can also rewrite the correct variation for the full determinant, $\nu = 0$,

\begin{equation}
  \frac{\delta S_{D}}{\delta u} = \frac{1}{2} \Bigl [ \partial_x (\bar{\Psi}\Psi) + u\frac{\delta \partial_x(\bar{\Psi}\Psi )}{\delta u} \Bigr ]. \label{det22_}
\end{equation}

\noindent
As a summary, if the action of the Burgers equation is claimed to be a number, we have a constraint for field $u$ in Eq.~(\ref{det22_}). 

\section{\label{sec4}Local symmetry}

In this section we investigate local symmetries of the effective action (\ref{acteff}). We  consider the following local time reparametrization, $\beta = \beta(t)$,  $a$ and $b$ are constants,

\begin{eqnarray}
  &&\tilde{t} = \beta(t)^a, \label{local1}\\
  &&\tilde{x} = x\beta'(t)^b ,\\
  &&\tilde{u} = u\beta'(t)^{a-b} ,\\
  &&\tilde{\mu} =\mu\beta'(t)^{2b-a},\\
  &&\tilde{{\Psi}} = \Psi\beta'(t)^{b/2} ,\\
  &&\tilde{\bar{\Psi}} =  \bar{\Psi}\beta'(t)^{b/2}. \label{local2}
\end{eqnarray} 

\noindent
This translates to the following field variations,

\begin{eqnarray}
  &&\delta u = a\beta u_t +b\beta'xu_x+(a-b)\beta' u,
\label{eq:deltau} \\
  &&\delta \mu = a\beta \mu_t +b\beta'x\mu_x+(2b-a)\beta'\mu,\\
  &&\delta \Psi = a\beta \Psi_t+b\beta' x\Psi_x+\frac{b}{2}\beta'\Psi, \label{eq:deltapsi2}\\
  &&\delta \bar{\Psi} = a\beta\bar{\Psi}_t+b\beta'x \bar{\Psi}_x+\frac{b}{2}\beta'\bar{\Psi}.\label{eq:deltapsi}
\end{eqnarray}

\noindent
 The transformation is based on symmetries (\ref{detu})-(\ref{detpsi}) and (\ref{0detu})-(\ref{0detpsi}) where 
 $\alpha(t) = a\beta(t)$ and $\epsilon(t,x) = b\beta'(t)x$.  

We first apply variations (\ref{eq:deltau}), (\ref{eq:deltapsi2}), (\ref{eq:deltapsi}) to action (\ref{det2}) ($\nu = 0$). This leaves the action invariant, $\delta S_{D}= 0.$ It follows that for consistency reasons also variation of action (\ref{det20}) must be zero when $\nu=0$,

\begin{equation}
  \delta \bar{S}_{D} = -\frac{1}{2}\int dtdx \frac{\beta'}{u_x}\partial_x (bxu_x+(a-b) u) = 0.
\end{equation}

\noindent
The invariance requirement creates a constraint for field $u$,

\begin{equation}
  bxu_x+(a-b) u = g(t), \label{constraint}
\end{equation}

\noindent
where $g(t)$ is a time dependent function.

Transformation (\ref{eq:deltau})-(\ref{eq:deltapsi}) for the action $S$ gives

\begin{equation}
  \delta S = \frac{3h-1}{2}b\int dtdxdy \beta' \mu(x)\kappa(x-y)\mu(y) 
- i C \int dtdx \mu+ i (h+1)\nu b \int dtdx \beta'\mu u_{xx},
\label{eq:final}
\end{equation}

\noindent
where $h=(b-a)/b$.  Here we have used constraint (\ref{constraint}) with $g(t) = C/\beta''$ and $C$ is a constant.  The pump term is zero when $h=1/3$. By setting $h = -1$ (or $\nu = 0$) the viscosity term is zero. These situations are explained in more detail in Ref.~\onlinecite{frisch}. The second integral is also zero based on the conservation of the center-of-mass motion, Ref.~(\onlinecite{ivaskevich}). 

Constraint (\ref{constraint}) defines a a solution for field $u$,

\begin{equation}
  u(t,x) = \tilde{g}(t)x^h-\frac{C}{bh\beta''(t)},
\end{equation}

\noindent
where $\tilde{g}(t)$ is a time dependent function. Variation of $u$ gets the form

\begin{equation}
 \delta u(t,x) = a\beta(t) u_t(t,x) +C \frac{\beta'(t)}{\beta''(t)}.
\end{equation}

Hence we have shown that in the limit of $\nu = 0$, symmetry transformation (\ref{local1})-(\ref{local2}) is a local symmetry for the effective action (\ref{acteff}). 

\

Our interpretation of this symmetry is, as it is explained in Ref.~\onlinecite{frisch}, that in the limit of infinite Reynolds numbers, $\nu \to 0$, all the symmetries of the Navier-Stokes equation are restored in a statistical sense The condition  $\tilde{u}(\tilde{x}) = u(x)$  is  also known as a Kolmogorov's second law of self-similarity, Ref.~\onlinecite{frisch}.

\section{\label{sec4}Conclusions}

In this paper we have studied local symmetries of the Burgers equation. By using a path integral formalism and the exact symmetries of the Jacobian we calculated the determinant. We then investigated a local symmetry of the effective action which relates to Kolmogorov's second law of self-similarity. We found an anomaly for the Burgers turbulence which shows as a constraint for the velocity field.

We thank A. Niemi for discussions.

\end{document}